\documentclass[twocolumn,aps,superscriptaddress,nofootinbib]{revtex4-1}
\usepackage{latexsym,amssymb,amsmath,epsfig,bm,psfrag}
\usepackage{color}
\usepackage[bookmarksnumbered,bookmarksopen,colorlinks,citecolor=red,linkcolor=blue]{hyperref}
\usepackage{mathrsfs}
\usepackage{times}
\usepackage{ulem}

\hypersetup{
    colorlinks=true,
    linkcolor=blue,
    filecolor=magenta,
    urlcolor=cyan,
    pdftitle={Sharelatex Example},
    bookmarks=true,
}

\newcommand{\bp}{{\bf p}}

\renewcommand\d{\delta}

\newcommand\g{\gamma}

\newcommand\m{\mu}
\newcommand\n{\nu}

\newcommand\p{\pi}
\newcommand\h{\theta}

\newcommand{\eq}[1]{Eq.~(\ref{#1})}

\newcommand{\non}{\nonumber\\}
\newcommand\pt{\partial}

\newcommand{\bx}{{\bf x}}

\renewcommand{\part}{{\rm part}}

\begin{document}

\title{Dynamical evolution of magnetic field in the pre-equilibrium quark-gluon plasma}
\author{Li Yan}\email{cliyan@fudan.edu.cn}
\affiliation{Institute of Modern Physics, Fudan University, Shanghai 200433, China}
\affiliation{Key Laboratory of Nuclear Physics and Ion-beam Application (MOE), Fudan University, Shanghai 200433, China}
\author{Xu-Guang Huang}
\email{huangxuguang@fudan.edu.cn}
\affiliation{Physics Department and Center for Particle Physics and Field Theory, Fudan University, Shanghai 200438, China}
\affiliation{Key Laboratory of Nuclear Physics and Ion-beam Application (MOE), Fudan University, Shanghai 200433, China}

\begin{abstract}
High-energy heavy-ion collisions generate extremely strong magnetic field which plays a key role in %can induce 
a number of novel quantum phenomena in %the produced 
quark-gluon plasma (QGP), %. A remarkable example is 
such as the chiral magnetic effect (CME). % which is under intensive experimental search at RHIC and LHC. 
%In particular, a quantitative understanding of these phenomena %observables for CME, %the knowledge of 
%relies on the knowledge of the time evolution of the magnetic field, %is essential 
%which, however, remains undetermined. 
However, due to the complexity in theoretical modellings of the coupled electromagnetic fields and the QGP system, especially in the pre-equilibrium stages, the lifetime %dynamical evolution 
of the magnetic field in the QGP medium remains undetermined.  We establish, for the first time, a kinetic framework to study the dynamical decay %evolution 
of the magnetic field in the early stages of a weakly coupled QGP by solving the coupled Boltzmann and Maxwell equations. We find that at late times a magnetohydrodynamical description of the coupled system emerges. With respect to realistic collisions at RHIC and the LHC, we estimate the residual strength of the magnetic field in the QGP when the system start to evolve hydrodynamically. 

%show that the induction %Faraday 
%effect in the pre-equlibrium QGP delays the decay of the magnetic field. 
%The dynamic evolution of the electric field is also studied. \blue{Our results pave the way to further quantitative studies of the CME and other electromagnetic field induced effects in heavy-ion collisions.}
\end{abstract}
\maketitle

{\it Introduction.}--- %{\color{blue} (Needs revision: 1/ EM fields and weakly coupled QGP. 2/ Realistic B field decay in the pre-equilibrium stage. 3/ Background, CME, heavy-flavor v1, etc. 4/ We focus on the very early-time stages: most B field decay, elastic collisions dominate, still far away from equilibrium. etc. 5/ Onset of MHD. 6/ Wide application of the coupling between EB and plasma.)} 
Quantum chromodynamics (QCD), the modern theory of strong interaction, predicts that at a sufficiently high temperature, %higher than $T_c\sim 200$ MeV 
quarks and gluons are liberated from hadrons and form a new state of matter --- the quark-gluon plasma (QGP). Creating and investigating QGP is the main purpose of BNL Relativistic Heavy Ion Collider (RHIC) and one of the main purposes of CERN Large Hadron Collider (LHC). In these experiments, due to the relativistic motion of ions and the smallness of the colliding systems, extremely strong magnetic fields %(with peak strength $|eB|\sim m_\p^2$ at RHIC and $|eB|\sim 10\,m_\p^2$ at LHC where $m_\pi$ is the pion mass)
(with peak strength $|eB| \sim 10^{19}$ Gauss at RHIC and $\sim 10^{20}$ Gauss at the LHC) are generated~\cite{Skokov:2009qp,Bzdak:2011yy,Voronyuk:2011jd,Deng:2012pc,Bloczynski:2012en}. Strong magnetic fields can significantly influence QGP and drive a number of interesting quantum phenomena~\cite{Kharzeev:2007jp,Fukushima:2008xe,Burnier:2011bf,Tuchin:2010vs,Basar:2012bp,Bali:2013owa,Tuchin:2014pka,Cao:2015cka,Fukushima:2015wck,Guo:2015nsa,Das:2016cwd,Sadooghi:2016jyf,Fukushima:2016vix}. One famous example is the chiral magnetic effect (CME) which
%When coupled with chiral anomaly, these magnetic fields
provides a feasible means to monitor the topological fluctuations of QCD in the high-temperature environments~\cite{Fukushima:2008xe,Kharzeev:2007jp}. %Although the data from RHIC and LHC show features consistent with the expectation of CME, there are severe background effects which strongly mask the desired signal~\cite{dd}. To disentangle the CME and the background effects, it is crucial to perform realistic simulations of the development of CME and make comparison with the experimental data. 
However, theoretical studies on such phenomena suffer from considerable uncertainties due to the lack of knowledge on the dynamical evolution of the magnetic fields in QGP. %Similar difficulties also appear in studies of other magnetic-field-induced effects, e.g., the chiral magnetic waves~\cite{dd} and the splitting of directed flow between $D^0$ and $\bar{D}^0$~\cite{dd}.
Especially, the lifetime of these fields, which essentially relies on how they decay along with the QGP evolution, remains unknown.

%The magnetic fields at the colliding moment %(i.e. the initial time) can be well estimated by computations based on well-established models like HIJING and UrQMD~\cite{dd}.  
Shortly after the nucleus-nucleus collisions, the created magnetic fields start to decay drastically due to the departure of the Lorentz contracted nuclei from the colliding zone, %Lorentz contraction of the colliding nucleus, 
inducing strong electric fields in the QGP. %The subsequent evolution of the electromagnetic (EM) fields in QGP remains largely unknown due to the difficulty in handling the coupled evolution of EM fields and QGP. 
It is a theoretical challenge to consistently describe the subsequent evolution of the coupled electromagnetic (EM) fields and QGP. Efforts have been provoked assuming a fully thermalized QGP with charge conductivity of the QGP as input, either solving Maxwell equations or based on magnetohydrodynamics (MHD)~\cite{McLerran:2013hla,Tuchin:2013apa,Li:2016tel,Gursoy:2014aka,Gursoy:2018yai,Inghirami:2016iru,Roy:2015kma,Roy:2017yvg}, indicating that induction in a QGP fluid %Faraday effect 
indeed slows down the decay of the magnetic fields. However, 
the {\it dominant} effects on the decay of magnetic fields from the very early stages, in which the magnetic fields decay most violently and the QGP is highly out-of-equilibrium, have not been discussed so far. 

It is the purpose of this Letter to perform, for the first time, a consistent computation of the dynamical evolution of EM fields in the pre-equilibrium QGP. We will establish a framework that based on a kinetic description for a weakly-coupled QGP system, i.e., assuming the strong coupling constant $\alpha_s$ to be small, and the Maxwell equations for the evolution of the EM fields. Since the kinetic description has been applied extensively to analyzing the onset of hydrodynamics in QGP ~\cite{Kurkela:2015qoa,Kurkela:2018vqr,Berges:2020fwq}, it is also our interest to investigate the emergence of MHD from the QCD plasma, in the presence of dynamically evolving EM fields.

%that couples the Boltzmann equation with the Coulomb-Lorentz force encoded describing the bulk evolution of QGP and the Maxwell equations describing the dynamical evolution of the EM fields. We will also examine how the MHD behavior is emerged during the thermalization procedure along with the evolution of EM fields.

{\it Methodology.}--- We start from the coupled equations that consist of the Maxwell equations which determine the dynamical evolution of the EM fields and a Boltzmann equation which determines the evolution of QGP,
%{\color{blue}( relation to MHD? )}
\begin{subequations}
\label{eq:eq0}
\begin{eqnarray}
\label{maxwell}
&\pt_\m F^{\m\n}=j_{\rm ex}^\n + j_{\rm ind}^\n\,,& \\
\label{boltzmann}
&[p^\m\pt_\m+e Q_a p_\m F^{\m\n}\pt_{p^\n}] f_a(t,\bx,\bp)={\cal C}[f_a]\,.&
\end{eqnarray}
\end{subequations}
In these equations, $F^{\m\n}=\pt^\m A^\n-\pt^\n A^\m$ is the EM-field strength tensor %,  $e>0$ is the the absolute value of electron charge, $Q_a$ is the charge number of parton of type $a=q,\bar{q}, g$ (i.e., quark, anti-quark, and gluon),
and $f_a$ is the distribution function of QGP constituents, viz. $a=q, \bar{q}$, or $g$ for quark, anti-quark or gluon, with $Q_a$ being the corresponding charge number. %In Eqs.~(\ref{eq:eq0}), 

The couplings between the EM fields and QGP are formulated in Eqs.~(\ref{eq:eq0}) in a complex way. For the EM fields, it is self-consistently introduced through the induced charge current,
\begin{eqnarray}
\label{eq:current}
j_{\rm ind}^\mu = e\sum Q_q\int\frac{d^3\bp}{(2\p)^3 }\frac{p^\mu}{E_q} h_q(t,{\bf x}, {\bf p})\,, %[f_q(t,z,\bp)-f_{\bar q}(t,z,\bp)],
\end{eqnarray}
where we have defined the splitting distribution $h_q \equiv f_q - f_{\bar q}$ %= \delta f_q - \delta f_{\bar q}$
to characterize the deviation between quark and anti-quark distributions due to the presence of EM fields. The summation in \eq{eq:current} %the induced charged current 
is over spin, flavor ($N_f=2$) and color ($N_c=3$). In addition to $j_{\rm ind}^\mu$, in heavy-ion collisions there is also an external charged current, $j_{\rm ex}^\mu$, generated by the moving spectators of the colliding nucleus, whose evolution is independent of the QGP medium. On the other hand, for quarks, the couplings are represented via the Coulomb-Lorentz force in Eq.~(\ref{boltzmann}). Note, however, gluons do not directly couple to the EM fields since they are electrically neutral.

%As being described by Eq.(\ref{eq:eq0}), quarks experience electromagnetic forces in addition to scatterings among quarks, anti-quarks and gluons determined by QCD, while gluons are not directly affect by the EM fields.
Unlike electromagnetic plasmas, scatterings in QGP among quarks, anti-quarks and gluons are dominated by QCD interactions, which are much stronger than the electromagnetic forces. %Lorentz force. 
Therefore, response in the quark and anti-quark distribution functions to EM fields can be treated as small corrections to the background distributions,
\begin{eqnarray}
\label{eq:deltaf}
f_{q}=\bar{f}_{q}+\d f_{q},\qquad f_{\bar{q}}=\bar{f}_{\bar{q}}+\d f_{\bar{q}}\,.
\end{eqnarray}
As a consequence of the smallness of $\d f_{q/\bar q}$, one could treat the background distributions of quark and anti-quark $\bar f_{q/{\bar q}}$ as being entirely determined by QCD interactions, so they are solutions to the Boltzmann equations: %without couplings to the EM fields:
\begin{eqnarray}
\label{boltz-fbar}
p^\mu \partial_\mu \bar f_{q/\bar q} = {\cal C}[f_{q/\bar q},f_g].
\end{eqnarray}
Due to the charge conjugation symmetry of QCD, one has $\bar f_q=\bar f_{\bar q}$ and whence $h_q = \delta f_q - \delta f_{\bar q}$.

%{\color{blue}(This following paragraph need to be written carefully: 1/ explain $|eB|\ll \Lambda_c^2$. 2/ when system approaches local equilibrium, $|eB|\ll \Lambda_c^2$ is violated, so that MHD. 3/ Can MHD be treated as the subsequent process after Boltzmann?) }
We now examine the condition $\delta f_{q/\bar q}\ll \bar f_{q/\bar{q}}$.  %We first notice that, 
By substituting Eq.~(\ref{eq:deltaf}) into  Eq.~(\ref{boltzmann}), one obtains an equation for $\delta f_{q/\bar q}$,
\begin{eqnarray}
p^\mu \partial_\mu \d f_{q/\bar{q}} + eQ_{q/\bar q} p_\mu F^{\m\n} \partial_{p^\n} (\bar f_{q/\bar{q}} + \d f_{q/\bar{q}}) = {\cal C}'[\d f_{q/\bar{q}}]\,,\,\,\,\,\,\,\,
\end{eqnarray}
where the collision kernel is deduced from the QCD interactions and is linear in $\d f_{q/\bar q}$.
The left-hand side of the equation reads parameterically : $\delta f_{q/\bar q} \Lambda_c^2 + \bar f_{q/\bar{q}}|eB|$, with $\Lambda_c$ a characteristic energy scale %and $L$ a length scale, respectively, %of quarks in QGP 
introduced from the kinematic term. To the leading order in $\alpha_s$, the collision term is further suppressed by $\alpha_s^2$, giving rise to $\alpha_s^2 \Lambda_c^2 \d f_{q/\bar{q}}$. %\footnote{More precisely, the suppression factor from collisions is $\alpha_s^2 L$, with $L$ the Coulomb logarithm. In total, the leading order contribution from the collision term linear in $\delta f_{q/\bar q}$ is parameterically $\alpha_s^2 L |eB|$.{\color{blue}(what is the inelastic and NLO expression?)}}.
Therefore, one finds $\delta f_{q/\bar q}\sim \bar f_{q/\bar{q}}|eB| /\Lambda_c^2$, from which $\delta f_{q/\bar q}\ll \bar f_{q/\bar q}$ translates into the condition $|eB|\ll \Lambda_c^2$. %This relation is guaranteed in heavy-ion collisions at the top energies of RHIC and the LHC.  Especially, r
Recall that at the top energies of RHIC and the LHC, at initial time the external magnetic field is estimated as $|eB|\sim m_\pi^2$ with $m_\pi$ the pion mass, while the only %both the 
characteristic energy scale $\Lambda_c$ %and the length scale of the system $L$ are 
in QGP is determined by the saturation energy, which is $Q_s\sim 1$ GeV. %, such that $\Lambda_c/L\sim Q_s^2$. 
As the system expands, both the magnetic field and the energy scale of the QGP medium decrease, but %{\it a priori} we claim that 
the condition $|eB|\ll \Lambda_c^2$ remains valid for a sufficiently long time, as will be shown later. Eventually, the condition breaks down when the system approaches local thermal equilibrium and becomes describable by MHD. For such a system, multiple energy scales emerge, including a typical hard energy scale corresponding to temperature $T$ and a soft energy scale corresponding to spatial gradient $\nabla$, %while the length scale is inversely proportional to spatial gradient, 
and $|eB| \gg  T\nabla $~\cite{Landau:712712,Hernandez:2017mch}. %Note that, for system close to local equilibrium, neglecting the collisions would effectively give rise to infinite conductivity.}
%\redh{[XGH: Initially, the system is characterized by one scale $Q_s$. When the system evolves towards thermalization, there would emerge different scales: the hard scale $T$, the soft scale for gluon distribution $\alpha_s T$, the screening scale for charge carriers $e T$ which is roughly the scale of $\nabla$ because the screening is due to the charge redistribution. The MHD applies when $|eB|$ is much larger than the screening scale, $|eB|\gg\alpha T^2\sim\nabla^2$ or $|B|\gg eT^2\sim T\nabla$.]}

Given $\delta f_{q/\bar q} \ll \bar f_{q/\bar{q}}$, for each flavor, %if one defines the splitting distribution $h_q \equiv f_q - f_{\bar q} = \delta f_q - \delta f_{\bar q}$, which characterizes deviations of quark and anti-quark distirbutions due to the effect of EM fields,
the splitting distribution can be shown approximately satisfying a collisionless Boltzmann-Vlasov equation,
\begin{align}
\label{eq:hq}
p^\mu\partial_\mu h_q + 2 eQ_q p_\mu F^{\mu\nu} \partial_{p^\nu} \bar f_q = 0\,, %{\cal C}[f_q] - {\cal C}[f_{\bar q}]\,.
%&- eQ_qp_\mu F^{\mu\nu} \partial_{p^\nu}(\delta f_q + \delta f_{\bar q})\non
%&+({\cal C}[f_q] - {\cal C}[f_{\bar q}] )\,.
\end{align}
where the Coulomb-Lorentz force %The second term in the left hand side of Eq.(\ref{eq:hq})
is dominated by the leading order mean-field effect, i.e., the drifting of the background quarks and anti-quarks in the EM fields. %\footnote{
%Deviations of the quarks and anti-quarks also drift in EM fields, which is apparently suppressed by an extra factor of $\alpha_{\rm EM}$, thus can be neglected in general.
%}.
With regards to \eq{eq:hq}, the charge carriers of the system that are coupled to the dynamical evolution of the EM fields are effectively the splittings between quarks and anti-quarks. For the leading order analysis, collisions are neglected as they are suppressed by $\alpha_s^2$ in comparison to the Coulomb-Lorentz force. In an alternative aspect, the collisionless approximation is also supported by the fact that the frequency of the charge carriers, $\sim \Lambda_c$, is much greater than that from collisions, which is $\sim\alpha_s^2 |eB|/\Lambda_c$~\cite{Lifshitz:99987}. Note that the collisionless approximation should be recognized as a consequence subject to the condition $|eB|\ll \Lambda_c^2$, namely, when the system is far away from equilibrium. For system close to local equilibrium, neglecting the collisions would give rise to infinite conductivity in QGP, which is apparently non-physical.  Accordingly, while the background distributions of quarks and anti-quarks can be solved independently from Eq.~(\ref{boltz-fbar}), the couplings between the EM fields and the pre-equilibrium evolution of QGP are characterized by Eq.~(\ref{maxwell}) and Eq.~(\ref{eq:hq}).

In high-energy heavy-ion collisions, in the absence of the EM fields, the pre-equilibrium evolution of QGP can be well approximated with respect to the Bjorken symmetry~\cite{Bjorken:1982qr}. That is to say, the background distribution of quarks and anti-quarks are boost invariant along the beam axis (we refer to as the $z$-axis) and translationally invariant in the transverse plane (${\bf x}_\perp$). Corresponding to realistic heavy-ion collisions, this approximation applies to the ${\bf x}_\perp\sim 0$ region in collisions that are not too peripheral. %Thus one can solve the background distrbution $\bar f_q(t, z, {\bf p})$ as the input for the Lorentz force in Eq.~(\ref{eq:hq}).

For the QCD interactions, we consider 2-to-2 elastic scatterings among massless up-, down-quarks and gluons: $q\bar{q}\leftrightarrow gg, qg\leftrightarrow qg, \bar{q}g\leftrightarrow \bar{q}g, gg\leftrightarrow gg$, at small angle approximation~\cite{Blaizot:2014jna}. We are allowed to vary the strong coupling constant in these scatterings when solving the equation, for which we shall choose as $\alpha_s\sim 0.2$. %, so the Boltzmann equation reduces to a Fokker-Planck equation.
%\redl{\sout{To avoid the exhibited flavor SU(3) symmetry among light quarks, which results in effectively an charge neutral QGP, we take $N_f=2$.}} %QGP thermalizes even without inelastic collisions, albeit after a rather long time. 
The inelastic collisions are not included since, at the very early stages as we are focusing on, they are expected subdominant~\cite{Baier:2000sb}, which thereby would not significantly change the results we find in the current work. Nevertheless, the role of inelastic collisions on the EM fields is an interesting topic to investigate, which we leave for future studies. More details on solving the background quark and anti-quark distributions can be found in Refs.~\cite{Blaizot:2017lht,Tanji:2017suk}, and Ref.~\cite{Churchill:2020uvk} for a recent application to the computations of electromagnetic probes.

In heavy-ion collisions, the magnetic fields on average are perpendicular to the reaction plane, which we set to be along the $y$ direction. In the reaction plane, there exist electric fields which are induced partly by the decay of the magnetic field and partly by the moving spectators. In accordance with this geometrical configuration, we choose the following gauge potential,
\begin{eqnarray}
\label{potential}
A^\mu=A_{\rm ex}^\mu + A_{\rm ind}^\mu = (0, A_{\rm ex}^x(t,z)+A_{\rm ind}^x(t,z), 0, 0),
\end{eqnarray}
which satisfies the Lorentz gauge condition $\partial_\mu A^\mu =0$.
Correspondingly, the independent EM-field components are given by
\begin{eqnarray}
\label{fields}
E^x(t,z) = -\partial_t A^x(t,z)\,,\quad
B^y(t,z) = \partial_z A^x(t,z) \,.
\end{eqnarray}
 In \eq{fields}, translational invariance in the transverse plane is implicitly assumed in analogous to the background QGP, which again makes our calculation mostly applicable at the central region of the transverse plane. %, $\bx_\perp\sim\bf 0$, away from which the transverse inhomogeneity may not be neglected.
%
%The time change of the magnetic field induces a Faraday current in QGP parallel to the reaction plane. We approximate this current by considering only its $x$ component, i.e., the component along the impact parameter direction. \blue{As we have checked, the $z$ component of the Faraday current is indeed smaller than its $x$ component.} Thus, we consider the following gauge potential for the EM fields,
%\begin{eqnarray}
%\label{potential}
%A^\mu=(0, A^x(t,z), 0, 0),
%\end{eqnarray}
%where the translational invariance in the transverse plane is implicitly assumed. This assumption significantly simplify the computation but makes our calculation mostly applicable at the central region of the transverse plane, $\bx_\perp\sim\bf 0$, away from which the transverse inhomogeneity may not be neglected.

The external EM fields generated by the spectator nucleons %generate currents along $z$ direction which create the external EM fields. These external EM fields
have been extensively studied. Near $\bx_\perp\sim\bf 0$ they can be well expressed via effectively a Lienard-Wiechert potential~\cite{Huang:2015oca,Hattori:2016emy}
\begin{eqnarray}
\label{backgroundfields}
A_{\rm ex}^x(t,z)&=&A_0\bigg\{\frac{z+vt}{[b^2/4+\g^2(z+vt)^2]^{1/2}}\non
&&+\;\frac{z-vt}{[b^2/4+\g^2(z-vt)^2]^{1/2}}\bigg\},
\end{eqnarray}
where $b$ is the impact parameter, $v=\sqrt{1-(2m/\sqrt{s})^2}$ is the velocity of the nucleus and $\g=1/\sqrt{1-v^2}$. $A_0$ is a constant parameter to specify the colliding systems, which depends on the atomic number of the colliding nucleus. The external EM fields evolve independently and satisfy by themselves the Maxwell equation, corresponding to which is the external charged current $j_{\rm ex}^\mu$. Accordingly, the Maxwell equation for the induced fields reduces to,
\begin{eqnarray}
\label{wave}
(\pt_t^2-\pt_z^2)A_{\rm ind}^x=-\pt_t E_{\rm ind}^x-\pt_z B_{\rm ind}^y=j_{\rm ind}^x\,,
\end{eqnarray}
with $j_{\rm ind}^x$ being given in Eq.~(\ref{eq:current}).

{\it Results.}---  With the above setup, we are able to solve the time evolution of both the QGP and the EM fields for given initial conditions. We take an initial condition for the gluon and quark distribution functions inspired by the saturation physics~\cite{Romatschke:2003ms}: 
\begin{eqnarray}
f_{g/q/{\bar q}}(t_0,z=0,\bp)=f_{g/q/{\bar q}}^{(0)}\,\h \left(1-\frac{\sqrt{p_z^2\xi^2+\bp^2_\perp}}{Q_s}\right).\,\,\,\,\,\,\,
\end{eqnarray}
Parameters in the initial condition will be taken according to the same strategy as in Ref.~\cite{Churchill:2020uvk}, namely, to be determined with respect to the multiplicity yields in realistic heavy-ion collisions. For instance, for the parameter $\xi$ which initializes the QGP out-of-equilibrium, we only consider the case $\xi= 1.4$, accounting for the effects of attractor solutions (cf. Ref.~\cite{Romatschke:2017vte} and Ref.~\cite{Shen:2020mgh} for a recent review).  %Since we are considering massless quarks, t
%The saturation scale $Q_s$ in the initial distributions provides the only scale in our formulation, so that time and spatial dependence can be expressed in unit of $Q_s^{-1}$. 
We choose the saturation scale $Q_s=1$ GeV, and accordingly we set $t_0 = 1\, Q_s^{-1}$ $\sim 0.2$ fm/c as the initial time of the system evolution. The constants $f_{q/\bar q}^{(0)}$ and $f_g^{(0)}$ specify the occupation of quarks and gluons at initial time. Except the constraint $f_{q/\bar q}^{(0)}\le1$ from Pauli exclusion principle, initial quark occupation should be further determined by the study of quark production before the kinetic regime through, e.g., the Schwinger mechanism in classical gluon fields~\cite{Gyulassy:1986jq,Nayak:2005pf}. %little is known about the initial quark occupation. %{\color{blue}(discussion on the possible initial quark occupation.)} 
In this work, we consider two limiting cases: Without initial quarks with $f_{q/\bar q}^{(0)}=0$ and quarks are fully populated initially with $f_{q/\bar q}^{(0)}=1$. %Accordingly, by adjusting $f_g^{(0)}$, one is allowed to fix the total entropy production in the QGP to reproduce the experimentally measured multiplicity yield for a given colliding event~\cite{Churchill:2020uvk}. 
The initial gluon occupation $f_g^{(0)}$ is then determined to reproduce the multiplicity yields in the 20-30\% centrality class of AuAu collisions at RHIC with $\sqrt{s_{NN}}=200$ GeV~\cite{Adler:2004zn} and PbPb collisions at LHC with $\sqrt{s_{NN}}=2.76$ TeV~\cite{Chatrchyan:2012mb}. %In summary, except initial quark occupations that we consider in the limiting cases, there are not free parameters in our simulations.

\begin{figure}
\begin{center}
\includegraphics[width=0.45\textwidth] {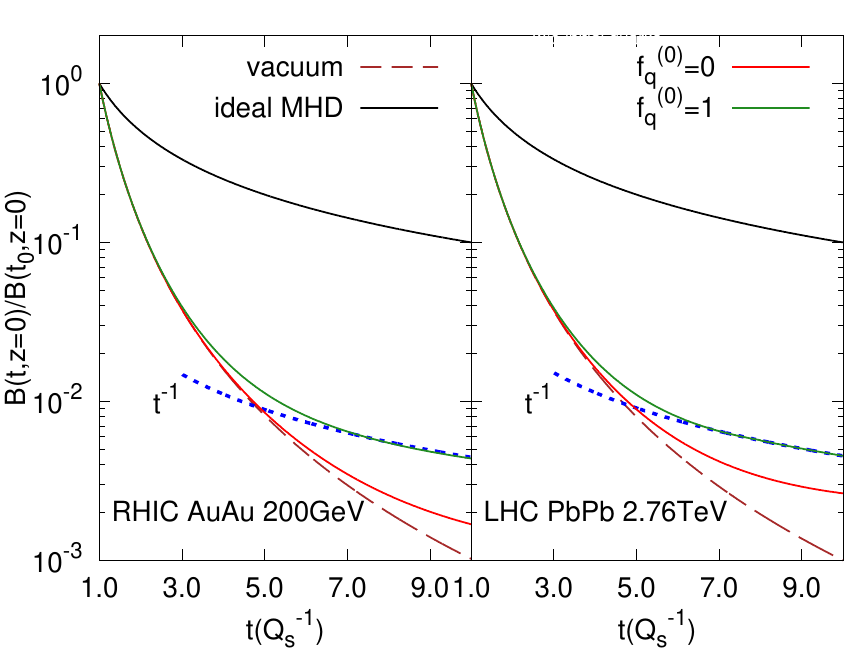}
\caption{ (Color online) Evolution of magnetic field at $z=0$ for the AuAu collisions at RHIC (left) and PbPb collisions at the LHC (right), relative to its initial strength at $t=t_0$. In comparison to the decay of magnetic field in vacuum (brown dashed lines), effects due to the QGP medium with initial quarks (red solid lines) and without initial quarks (green solid lines) are obvious. The expected $t^{-1}$-decay from ideal MHD is plotted as the black solid lines and blue dotted lines. 
\label{fig:evoB}
}
\end{center}
\end{figure}

With all these parameters given, we can solve the coupled differential equations numerically. Results shown in Fig.~\ref{fig:evoB} are the time evolution of the magnetic field relative to its initial value at $z=0$, i.e., the center in the QGP medium, for AuAu collisions at RHIC and PbPb collisions at the LHC. For comparison, we also present the results of two extreme scenarios in which analytical solutions exist. The first is the evolution of the magnetic field in vacuum, where the decay of magnetic field is entirely determined by the spectator nucleons moving relativisticly, resulting a time dependence $\sim t^{-3}$~\cite{Hattori:2016emy}. This is exactly the evolution of the external field in our study. The second corresponds to solutions of ideal MHD, with regard to an infinite electric conductivity $\sigma\to \infty$. The evolution of the magnetic field is then determined by the magnetic flux conservation in the conducting QGP. For the Bjorken expansion it is estimated as $\sim t^{-1}$~\cite{Deng:2012pc,Roy:2015kma}. For more realistic  heavy-ion collisions, as a consequence of the QGP evolving towards local equilibrium, the evolution of the magnetic field interpolates these two limits. For example, for the solution with respect to $f_q^{(0)}=1$ (green lines in Fig.~\ref{fig:evoB}), the  magnetic field starts following the vacuum solution $t^{-3}$, then tends to behave $\sim t^{-1}$ at late times. Deviations from the vacuum solutions are those generated from induction.

%The time evolution of the magnetic field reflects the evolution of QGP towards chemical equilibrium.
In analogous to the hydrodynamization process characterized by the evolution of energy density in a pre-equilibrium QGP~(cf. Ref.~\cite{Kurkela:2015qoa}),  the evolution of the magnetic field can be used to monitor the onset of MHD. However, the evolution of the magnetic field is more involved as it reflects not only the dynamical aspect of the QGP, but also depends on its chemical evolution. Initially, irrespective to quark occupation, QGP is dominated by gluon saturation, which leads to approximately a charge neutral medium that barely couples to the EM fields, so it explains the $t^{-3}$ behavior at early times. As the system evolves towards chemical equilibrium with quarks generated gradually via scatterings, the medium becomes more and more conducting and eventually evolves according to the MHD description. In particular, the $t^{-1}$-decay of the magnetic field is characteristic in ideal MHD in the Bjorken flow, %due to the conservation of the flux of the magnetic field, 
hence it is the final and the {\it slowest} decay pattern of the magnetic field. However, from our simulations the $t^{-1}$-decay of the magnetic field emerges much earlier than the applicability of MHD when the coupled system is still out of equilibrium, %for which the condition $|eB| \gg \Lambda_c^2 \sim T \nabla $ is required. 
a phenomenon that is also observed in the collisionless electromagnetic plasmas~\cite{Lifshitz:99987}. Of course, the condition $|eB| \gg T \nabla $ for local equilibrium and 
MHD should eventually be realized at later times, because $T\nabla\sim t^{-4/3}$ decays faster than the magnetic field.
%as %chemical equilibrium in QGP, nor
%the onset of ideal MHD.  {\color{blue} (Why? Effectively infinite conductivity from the neglecting of the collisions, and one thus have approximately the conservation of the magnetic field flux, but it is not the MHD since $|eB| \ll \Lambda_c^2$ is still valid. To understand the validity of the relation, we recall that the time dependence of the energy scale (such as temperature) in Bjorken expansion starts from $\Lambda_c^2\sim t^{-1/2}$ to $\Lambda_c^2\sim T\nabla \sim t^{-4/3}$.)} Nevertheless, it
It is also expected that chemical equilibrium in the QGP is more easily realized when quarks are populated initially, which explains in Fig.~\ref{fig:evoB} the slower decay of magnetic field in QGP with initial quarks than that without initial quarks.  

\begin{figure}
\begin{center}
\includegraphics[width=0.45\textwidth] {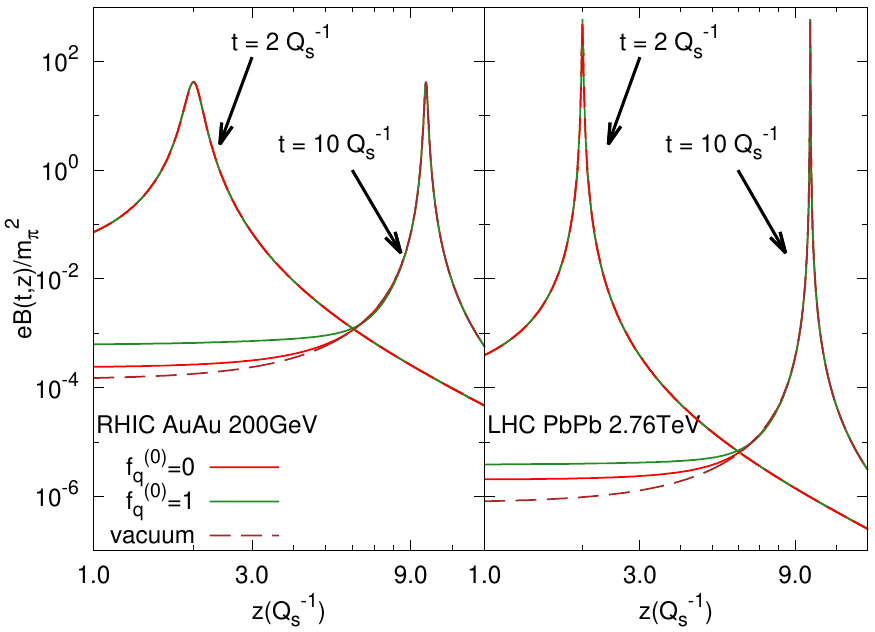}
\caption{ (Color online) Distribution of magnetic field along $z$-axis at $t=2\, Q_s^{-1}$ and $t=10\, Q_s^{-1}$, in units of $m_\pi^2$, for AuAu collisions at $\sqrt{s_{\rm NN}}=$200 GeV (left) and PbPb at $\sqrt{s_{\rm NN}}=$2.76 TeV (right). Numerical solutions with respect to QGP medium initialized with $f_q^{(0)}=0$ and  $f_q^{(0)}=1$ are plotted as red solid lines and green solid lines, respectively. For comparison, the distributions of magnetic field in vacuum are shown as brown dashed lines.
 \label{fig:disB}
}
\end{center}
\end{figure}

%\begin{itemize}
%\item Non-equilibrium QGP delays the decay of magnetic field, comparing to vacuum, but the effect is much weaker than that in an ideal MHD.
%\item On initial quark occupation.
%\item There are relatively more quarks at RHIC than the LHC, or to say, QGP at RHIC is more conducting.
%\item Due to the larger Lorentz factor at the LHC, magnetic fields drops much faster.
%\item Critical significance of the absolute value of the magnetic field strength: $(m_q/m_\pi)^2\sim 10^{-4}$ and
%$(m_e/m_\pi)^2\sim 10^{-5}$.
%\item On effective charge conductivity:  Farady current and Hall current.
%\item Heavy flavor $v_1$ puzzle: qualitatively agrees with the case with $f_q^{(0)}=1$.
%\item Isobar ...
%\item Why we expects more quarks initially, from the classical gluon fields.
%\end{itemize}

Although the effects of induction from the out-of-equilibrium QGP are comparable at RHIC and LHC, as being recognized from the relative decay in Fig.~\ref{fig:evoB}, %when QGP approaches local equilibrium,
the absolute strength of magnetic field is much stronger at RHIC. This is because the field strength relies, to a larger extent, on its external field component. Shown in Fig.~\ref{fig:disB}, are the magnetic fields along $z$-axis, plotted in units of pion mass square, at $t=2\, Q_s^{-1}$ and $10\,Q_s^{-1}$.
%$t=10\, Q_s^{-1}$ $\sim 2$ fm/c is taken the final time step of the pre-equilibrium evolution of the QGP. 
We emphasize that at $t=10\,Q_s^{-1}\sim 2$ fm/c, the QGP system is expected to evolve hydrodynamically, yet it is still out-of-equilibrium~\cite{Blaizot:2017lht,Kurkela:2018vqr}.
The peaks of these distributions point to the positions of the colliding nuclei at the respective instants. Again, induced magnetic field can be noticed as the difference between the vacuum solution and the solution with respect to QGP, which, as expected, only becomes significant at late times and in the spatial area where the QGP medium exists. %Even if initial quarks are fully presumed, the residual magnetic field at the final time scales only $|eB|\sim 5\times 10^{-6} m_\pi^2$ in PbPb collisions at the LHC, comparing to $|eB|\sim 10^{-3} m_\pi^2$ in AuAu collisions at RHIC. 
The critical scales in a QGP medium correspond to the mass of light quarks, $(m_q/m_\pi)^2\sim 10^{-4}$, and temperature $(T/m_\pi)^2\sim (\Lambda_{\rm QCD}/m_{\pi} )^2$. %and $(m_e/m_\pi)^2\sim 10^{-5}$, respectively,
In comparison to these scales, we found, by the time the QGP starts to evolve hydrodynamically, the residual strength of the magnetic field satisfies the hierarchy, $(m_q/m_\pi)^2\ll|eB|/m_\pi^2\ll(T/m_\pi)^2$ at RHIC, but can be negligible at the LHC. This is possibly the reason why CME has not been detected at the LHC experiments~\cite{Khachatryan:2016got}.

%{\color{blue}(Why we have many quarks from the classical gluon fields.)}

\begin{figure}
\begin{center}
\includegraphics[width=0.45\textwidth] {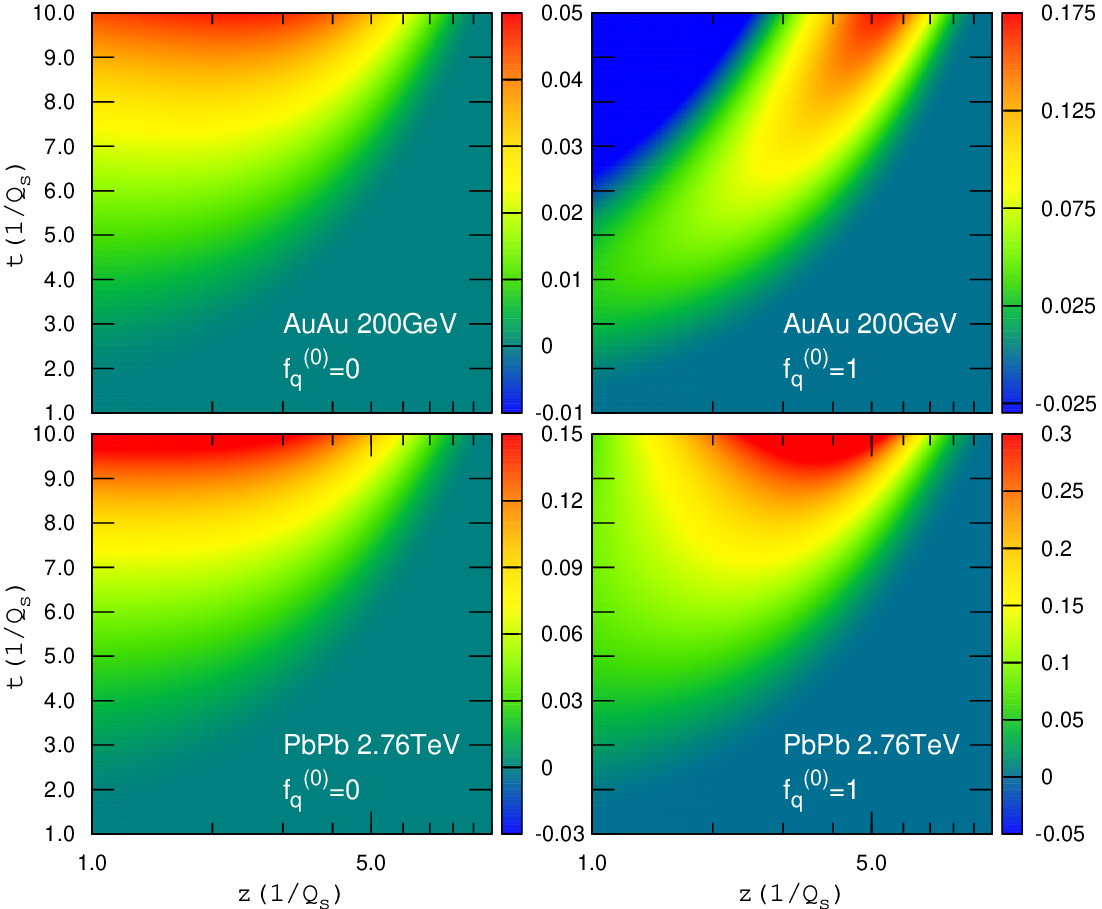}
\caption{ (Color online) The ratio %Effective electrical conductivity calculated by 
$j_{\rm ind}^x/E^x$ from AuAu collisions at $\sqrt{s_{\rm NN}}=$200 GeV (up) and PbPb at $\sqrt{s_{\rm NN}}=$2.76 TeV (down), with initial quarks (right) and without initial quarks (left), in unit of $Q_s$.
 \label{fig:jovere}
}
\end{center}
\end{figure}

Fig.~\ref{fig:jovere} presents the ratio $j_{\rm ind}^x/E^x$ at RHIC and the LHC collisions, in units of $Q_s=1$ GeV. The purpose of studying this ratio is two-fold. First, as a straightforward generalization of the electric conductivity to systems out-of-equilibrium, $j_{\rm ind}^x/E^x$ characterizes the conducting ability of the QGP during its pre-equilibrium stages. Note however, due to the influence of the magnetic fields on the QGP medium, one should not simply identify the ratio to the electric conductivity. Since the background QGP experiences Bjorken expansion, the ratio is nonzero only in the region with $z<t$. As shown in Fig.~\ref{fig:jovere}, the QGP becomes more conducting as it evolves towards equilibrium, reflecting the continuous quark and anti-quark production, in consistency with what we have learned before. 
%For a QGP close to equilibrium, it is expected from Lattice QCD that the conductivity should be proportional to temperature, $\sigma\sim 0.2 T/T_c$, which obviously decreases as the system cools down. %When quarks are populated initially, the QGP gets more easily conducting, which explains a factor of two and a factor of three increases of the ratio at RHIC and the LHC, respectively. 
Secondly, $j^x_{\rm ind}$ signifies the motion of charged particles in the reaction plane due to the presence of EM fields. It gives, in particular, one major origin of the charge dependent flow $v_1$ observed in experiments~\cite{Adam:2019wnk,Acharya:2019ijj}. Importantly, the induced current contains contributions from both the electric and the magnetic fields, but in opposite directions: $j_{\rm ind}^x \propto E^x - v_z B^y$. These distinct field components are relevant to the rapidity dependence of the observed $v_1$\footnote{
For instance, for the positively charged hadrons, the electric field can lead to $d v_1(y)/dy>0$, while the magnetic field tends to drive $dv_1(y)/dy<0$~\cite{Das:2016cwd}.
}, whose relative significance can be investigated in term of the sign of $j_{\rm ind}^x/E^x$. For all the cases we are considering, the induced current is dominated by the electric field component, thus $j_{\rm ind}^x/E^x>0$. However, an exception $j_{\rm ind}^x/E^x<0$ is observed in the central region at the RHIC AuAu collisions when initially quarks are fully populated (top-right panel). Here the dominant effect comes from the magnetic field. It is worth mentioning that, couplings to EM fields are more substantial for the transport of heavy-flavor quarks, leading to a more significant effect. Interestingly, the sign change of rapidity dependence in the heavy-flavor $v_1$ in experiments from RHIC to the LHC~\cite{Adam:2019wnk,Acharya:2019ijj}, is consistent to the switching dominance between the electric field and magnetic field observed here, with respect to the condition that initially more quarks are populated at RHIC.

{\it Summary.}--- We have presented the evolution of magnetic field in the pre-equilibrium stages of the QGP  for realistic heavy-ion collisions, from the numerical solutions to the coupled Boltzmann-Vlasov equation and Maxwell equations. %With parameters chosen according to the realistic collision events in experiments, we are able to study the evolution of the magnetic field, the residual strength the magnetic field when the QGP is hydrodynamized, and the relative significance of the electrical field and the magnetic field.
We showed that the electric induction in QGP slows down the decay of magnetic field and found that the residual strength of the magnetic field after pre-equilibrium evolution is still strong at RHIC in comparison to the critical scale given by light quark mass, but negligible at the LHC. We found that the evolution of the magnetic field approaches the description of an ideal MHD, namely, $t^{-1}$-decay, at later times, % when the background QGP starts to evolve hydrodynamically, 
although the applicability condition $|eB| \gg T \nabla $ for MHD has not been achieved. We also studied the ratio $j_{\rm ind}^x/E^x$, which is in general dominated by its electric field component. However, when the effect of magnetic field is enhanced such as in the RHIC collisions with initial-state quarks, the induced charged current follows the direction determined by the magnetic field. Our findings can provide initial input for MHD simulations for heavy-ion collisions and may also underlie the quantitative computations of the magnetic-field induced observables like the CME signal and the charge-dependent directed flow of heavy flavors.%This observation is particularly important when it applies to the evolution of heavy-flavor quarks. It is worth mentioning that, the induced charge current is the major force that pushes the heavy-flavor quarks moving in the reaction plane, hence the source of the observed charge-dependent $v_1$ in experiments. As one may examine, the observed sign change of the induced current $j^x_{\rm ind}$ in our simulations, is consistent to the sign change of the charge-dependent $v_1$ in experiments~\cite{}.

{\it Acknowledgments.}---
X.-G.~H. is supported by NSFC through Grant No.~12075061 and Shanghai NSF through Grant No.~20ZR1404100. L.~Y. is supported by the NSFC Grants through No.~11975079 and by Shanghai Pujiang Program (No. 19PJ1401400).

%--- Bibliography ---%
\bibliography{references}
%--- Bibliography ---%

\end{document}